\documentclass[graybox]{article}

\usepackage{graphicx}

\usepackage[svgnames]{xcolor}

\usepackage{natbib}
\bibpunct{(}{)}{;}{a}{}{,}  

\usepackage[stable]{footmisc}

\usepackage{hyperref}
\hypersetup{colorlinks,
            linktocpage,
            breaklinks,%
            citecolor=blue,%
            filecolor=blue,%
            linkcolor=blue,%
            urlcolor=blue,}

\usepackage[font=small,labelfont=bf]{caption}

\usepackage[tight,nooneline,FIGTOPCAP,bf]{subfigure}
\renewcommand{\thesubfigure}{\alph{subfigure}}
\makeatletter
    \renewcommand{\@thesubfigure}{(\thesubfigure)}  
    \renewcommand{\@@thesubfigure}{\thesubfigure}   
    \renewcommand{\p@subfigure}{\thefigure}         
\makeatother

\usepackage{fancyhdr}

\usepackage{combelow}
\usepackage{algorithm}
\usepackage{algpseudocode}
\usepackage{enumerate}
\usepackage{amssymb,amsmath}
\usepackage{moreverb}
\usepackage{listings}
\usepackage{multirow}
\usepackage{verbatim}
\usepackage{epsfig}
\usepackage{epstopdf}

\usepackage[leftcaption]{sidecap}

\usepackage{titlesec}
\titlelabel{\thetitle.\quad}

\renewcommand{\thefigure}{\thechapter.\arabic{figure}}
\makeindex             

\begin{document}

\title{The when and where of water in the history of the universe}
\author{Karla de Souza Torres$^{\rm 1}$, and Othon Cabo Winter$^{\rm 2}$\\
$^{1}${\em{CEFET-MG, Curvelo, Brazil; E-mail: karlchen79@gmail.com}}\\
$^{2}${\em{UNESP, Grupo de Din\^ amica Orbital \& Planetologia, Guaratinguet\'a, Brazil}}\\
{\em{E-mail: ocwinter@pq.cnpq.br}}}

\date{}
\maketitle

\begin{flushleft}
\textbf{Abstract}\\
\end{flushleft}

It is undeniable that  life as we know it depends on liquid water. It is difficult to imagine any biochemical machinery that does not require water. 
On Earth, life  adapts to the most diverse environments and, once established, it is very resilient. Considering that water is a common compound in 
the Universe, it seems possible (maybe even likely) that one day we will find life elsewhere in the universe. In this study, we review the main aspects 
of water as an essential compound for life: when it appeared since the Big Bang, and where it spread throughout the diverse cosmic sites. Then, 
we describe the strong relation between water and life, as we know it. \\

\begin{flushleft}
\end{flushleft}

\section{ Introduction. Why water is essential for life?}
\label{sec:1}

It is well known that liquid water has played the essential and undeniable role in the emergence, development, and maintenance of life on Earth. 
Two thirds of the Earth's surface is covered by water, however fresh water is most valuable as a resource for animals and plants. Thus, 
sustainability of our planet's fresh water reserves is an important issue as population numbers increase. Water accounts for 75\% of human
body mass and is the major constituent of organism fluids. All these facts indicate that water is one of the most important elements for
life on Earth. Thus, ``follow the water" has become a mantra of the science of astrobiology \citep{Irion2002}. 

Water is present on the surface of our planet at ambient temperatures and pressures in three different states: liquid, vapor and solid.
Water is also found everywhere in the universe: in the most distant galaxies, among the stars, on the Sun, on planets and their satellites 
and ring systems, in asteroids, and in comets. Water exhibits unique properties that make it extremely important for life as known on Earth.
First, it is the only substance on Earth that is abundant in liquid form at temperatures commonly found on the planet's surface. Second, it
is a superb solvent, implying that other substances can easily dissolve in it. Thus, water carries nutrients to cells and is used to wash
away the waste \citep{Lynden-Bell2010}. 

Water is formed from two very abundant elements in the universe: hydrogen, the most abundant element, and oxygen, the third (Helium is second)
most abundant element in the universe. 

In this chapter we discuss when did water appear after the Big Bang, where did it spread in the universe, what potential roles did it play in
the emergence of life? First we discuss the  physical and chemical characteristics, and then, turn to the history of water in the universe, its 
cosmic formation, and abundance. Next, we individually discuss the diverse cosmic sites, from distant galaxies to nearer stars and planets,
where water has been discovered. Finally, we describe the strong relation between water and life, as we know it, and formulate conclusions about
the great endeavor of determining the when and where of water in the universe's history.

\section{ What is water?}
\label{sec:2}

\subsection{Chemical properties of water}
\label{subsubsec:2-1}

Water molecule includes two hydrogen atoms and one oxygen atom, connected with a strong polar covalent bond. Water molecule are stable and 
can last for millions and even billions of years. Water forms in a chemical reaction of two molecules of hydrogen (H$_2$) with one molecule
of oxygen (O$_2$2), as shown in Eq. .1: 

\begin{eqnarray}
H_2 + H_2 + O_2 = H_2O + H_2O
\label{eq:01}
\end{eqnarray}

This process is one of the most exothermic among known chemical reactions, with released energy of 572 kJ/mol \citep{hanslmeier2010water}. 

\begin{figure}
  \centering
  \caption{Structure of the H$_2$O molecule, electrically polarized.}
  \includegraphics[width=0.7\textwidth]
    {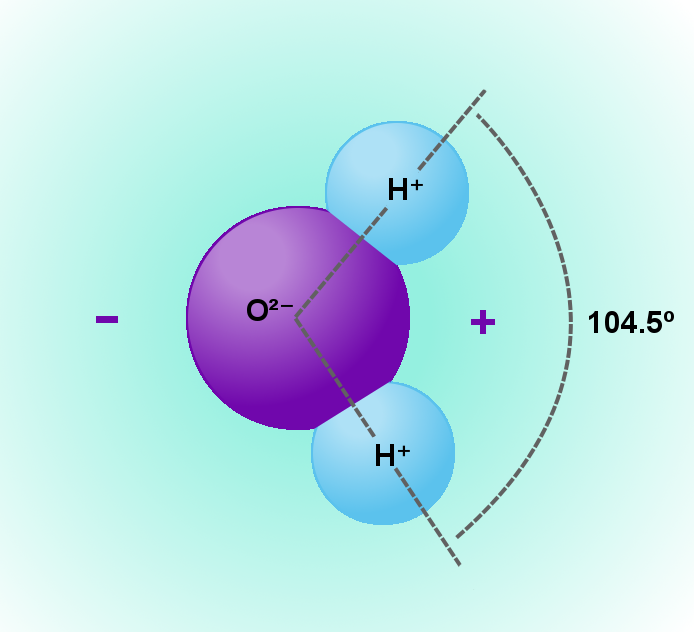}
  \label{watermolecule}       
\end{figure}

The water molecule is not linear but has a shape of a triangle (Fig\ref{watermolecule}). At the apex is the oxygen atom,
with two hydrogen atoms forming an angle of $104.5\,^{\circ}$. With six valence electrons, oxygen needs eight to fill its valence shell. 
Thus, it shares two electrons from the two hydrogen atoms, which become positively charged.  Since the small hydrogen atoms have weaker 
affinity for electrons than the large oxygen atom, the molecule is bent, and the two hydrogen atoms appear on the same side. Water is thus
classified as a polar molecule because of its polar covalent bonds and its bent shape \citep{Encrenaz2007searching}. 

The polarity of water explains why it is an excellent solvent. Water can induce temporary dipoles in even non-polar molecules, and 
interact differently with charged and polar substances. The polar molecules interact with the partially positive and negative ends of 
the water molecule, which results in the formation of a three-dimensional sphere of water molecules surrounding the solute. Water can 
thus solve and accumulate a variety of substances that are important for life. If the bonds were linear, then water would not be a strong solvent, which could affect the origin of life on Earth.

\subsection{Physical properties}
\label{subsubsec:2-2}
  
On Earth, water is the only substance that exists naturally in all three phases: gas, liquid, and solid. Its unique properties have allowed
life to be possible on Earth. Hanslmeier \citep{hanslmeier2010water} wrote: 

\begin{itemize}
\item  Liquid water is more dense than water ice. This is essential for life because ice always forms at the surface, protecting life bellow
it from freezing; 
\item  The pH of pure water is neutral, value of 7, which is neither acidic nor basic; 
\item  Water boils at $100\,^{\circ}\mathrm{C}$ and freezes at $0\,^{\circ}\mathrm{C}$ under normal pressure conditions.
\end{itemize}

Liquid water boils at a temperature at which vapor pressure reaches the environmental pressure around the liquid. The higher is the
environmental pressure, the higher is the temperature at which the liquid will boil. This temperature is known as the boiling point. At standard 
pressure\footnote{The standard pressure is the pressure at 1 bar ($100 kPa$, the current IUPAC - International Union of Pure and Applied 
Chemistry - definition). The standard reference conditions are: temperature $0\,^{\circ}\mathrm{C}$, pressure $100 kPa$. The standard
atmosphere (symbol: atm) is a unit of pressure defined as 101,325 Pa (1.01325 bar)}, 
water boils at $100\,^{\circ}\mathrm{C}$. The pressure on the top of Mount Everest is $260 Pa$, where the boiling point of water 
is $69\,^{\circ}\mathrm{C}$.

\begin{figure}
  \centering
  \caption{Log-lin pressure-temperature phase diagram of water. The Roman numerals indicate various ice phases. Credit: By Cmglee (original work), via Wikimedia Commons}
  \includegraphics[width=0.9\textwidth]
    {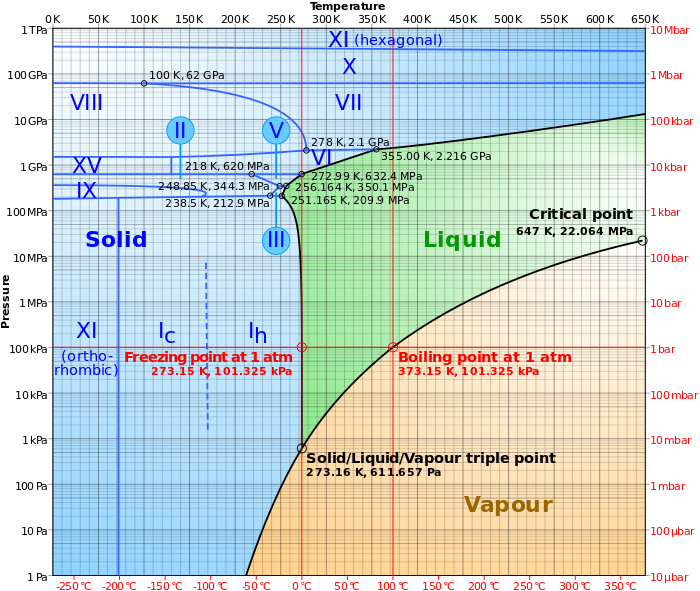}
  \label{waterphase}       
\end{figure}

The curves on the phase diagram shown in Fig\ref{waterphase} correspond to the boundaries between the different phases of water,
according to the temperature and pressure. The triple point, at the intersection of the three curves, indicates the pressure and 
temperature where water can coexist in all three phases. It is at a temperature of $0.01\,^{\circ}\mathrm{C}$ ($273.16 K$) and a 
pressure of $611.657 Pa$. At a low pressure of just $7.000 Pa$, water boils at $38.5\,^{\circ}\mathrm{C}$. This is about one order
of magnitude higher than the atmospheric pressure on Mars. Therefore, liquid water cannot exist on the Martian surface at present. 
Despite that, salty liquid water seems to flow from some steep, relatively warm slopes on the surface of Mars \citep{gough2016formation}. 
This is addressed in the Section \ref{subsec:4-5-3}.

\section{When did water appear?}
\label{sec:3}

The raw materials for producing water molecules are hydrogen and oxygen atoms. Here we discuss where these elements come from, and  how they were formed. 

Spectroscopic analysis of sunlight indicates that the Sun's photosphere is composed of hydrogen (74.9\%), helium (23.8\%), and other heavier elements \citep{Lodders2003, Lodders2003a}. Among the heavier elements, the most abundant is oxygen (around 1\% of the solar mass) \citep{Hansen2004}. The abundance of the metals is usually estimated considering not just the spectroscopy of the Sun's photosphere but also the measures from meteorites that are believed to contain the original Solar composition \citep{Piersanti2007}. 

However, our Sun was not created in the first generation of stars in the universe. It is only approximately 4.6 billion years old, while the universe is known to be approximately 13.7 billion years old. In fact, the chemical composition of our Sun was inherited from the interstellar medium, which was produced from previous generations of stars. Moreover the ingredients of the first stars were generated in the Big Bang nucleosynthesis. 

The simplest chemical element in nature is hydrogen. Through stellar evolutions and nuclear fusion hydrogen leads to other elements, including oxygen. In the current section, we present how, when, and where these elements were and are being created in the universe.

\subsection{Primordial nucleosynthesis}
\label{subsec:3-1}

After the Big Bang, the early universe was initially very dense and hot. It cooled down as it expanded, and the quark-gluon plasma gave origin
to neutrons and protons (and other hadrons, but in very small quantities). The universe continued to cool, and quickly (after 15 to 30 minutes) 
the nucleosynthesis ceased because it became too cold \citep{Rauscher2011}. The decay time of a free neutron is approximately 10 min. However,
before their decay, neutrons interact with protons forming deuterium nuclei. The deuterium obtains another neutron and form tritium, which in
turn absorbs a proton to form a $^4$helium. There is no stable element of mass 5 or 8. Therefore, it is generally not possible to have additional
nucleosynthesis via H + $^4$He = $^5$Li or $^4$He + $^4$He = $^8$Be; nevertheless, traces of one or two heavier elements, most notably $^7$lithium,
do form. Most of the matter was then hydrogen and $^4$helium, with a small amount of deuterium, and just traces of $^3$helium and $^7$lithium. 
Neutrons and protons started to form only after the first 1/1000th of a second from the Big Bang, when the temperatures dropped low enough. From
the first 1/100th of a second up to 3 or 4 min after the Big Bang, the abundances of the first very light atomic nuclei were defined. The ratio of
cosmic abundance today, expressed in terms of mass, is approximately 71\% of hydrogen, 28\% of helium, and 1\% for all the remaining elements. 
However, at most 4\% of this helium could be the result of burning hydrogen inside of stars since the beginning of the universe. Then, the initial 
ratio must have been approximately 24\% of helium and 76\% of hydrogen \citep{Rauscher2011}.

Therefore, the primordial stars in the universe formed from a gaseous mixture of hydrogen and helium, as well as a very few traces of some rare 
light elements, such as 7lithium, or isotopes such as deuterium or 3helium \citep{Karlsson2013}. They did not have any oxygen.

\subsection{Energy production in stars}
\label{subsec:3-2}

One of the most intense research areas in the early 20th century was the source of stellar energy. A seminal paper on the subject was written 
by Hans Bethe in 1939, entitled ``Energy Production in Stars", in which he presented two processes as being the main sources of stars' 
energy \citep{Bethe1939}. In 1967, Bethe received the Nobel Prize in Physics for his discovery. The first process is the proton-proton chain 
reaction (see fact box 1), which is the main source of energy for stars with the same or smaller mass than that of the Sun. However, the 
carbon-nitrogen-oxygen cycle (CNO; see fact box 2), which was also considered by \cite{vonWeizsacker1938}, is the one that dominates in more 
massive stars. 

It is interesting to note that the initial goal was to explain the source of energy of the stars, but these studies also showed how some light
chemical elements could have been generated. The studies of Bethe did not address the creation of heavy nuclei. This was studied later by 
Fed Hoyle \citep{Hoyle1946, Hoyle1954}. He showed that stars with advanced fusion stages were able to synthesize elements in the mass range 
from carbon and iron. The works of Hoyle are considered fundamental for the field of stellar nucleosynthesis \citep{Clayton1968, Clayton2008}.

\vspace{5mm}

\begin{figure}
\centering
\caption{The proton-proton chain reaction. CREDIT: Wikipedia}
  \includegraphics[width=0.6\textwidth]
    {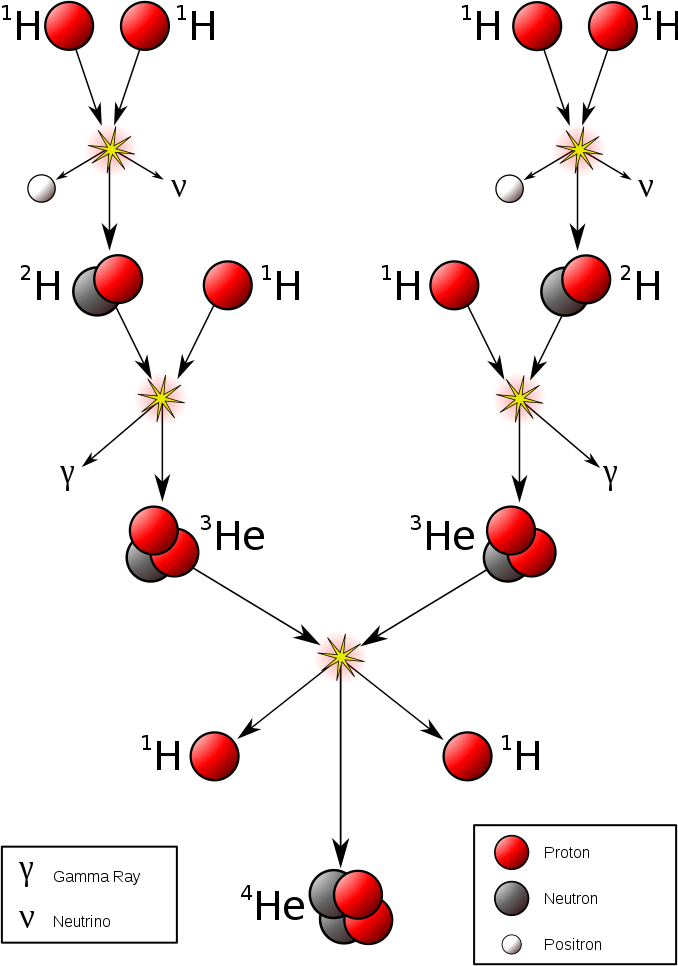}
  \label{proton-proton}      
\end{figure}

\fcolorbox{gray}{pink}{
\begin{minipage}[c]{.9\textwidth}
\begin{center}
\textbf{BOX 1: PROTON-PROTON CHAIN}
\end{center}
The fusion of hydrogen occurs primarily following a chain of reactions called proton-proton chain \citep{Wallerstein1997}:

\begin{eqnarray}
4 {^1H}  \longrightarrow  2  {^2H} + 2 e^+ + 2 \nu_e  \nonumber\\
2   {^1H} + 2  {^2H}  \longrightarrow  2  {^3He} + 2 \gamma \nonumber\\
2   {^3He}  \longrightarrow  {^4He} + 2  {^1H} 
\label{eq:02}
\end{eqnarray}

The overall reaction corresponds to the following equation: 

\begin{equation}
4 {^1H}  \longrightarrow  {^4He} + 2 e^{+} + 2 \nu_e + 2 \gamma
\label{eq:03}
\end{equation}

As illustrated in Figure \ref{proton-proton}, four nuclei of hydrogen (i.e., protons) collide in pairs of two. Each 
collision results in a nucleus of deuterium, positron, and neutrino. The positrons collide with electrons and become
annihilated, emitting gamma rays, whereas each nucleus of deuterium collides with a nucleus of hydrogen (proton) generating
a nucleus of $^3$He and emitting energy. In the last stage of the cycle, the two nuclei of $^3$He are fused forming a 
nucleus of helium ($^4$He) and two nuclei of hydrogen. In stars of the mass our Sun or smaller, the proton-proton chain 
is the dominating reaction. In the core of the Sun, the proton-proton chain occurs approximately $9.2 \times 10^{37}$ times 
per second, converting $3.7\times 10^{38}$ protons into helium nuclei \citep{Phillips1995}.

\label{box1}
\end{minipage}
}

\vspace{5mm}

\vspace{5mm}

\begin{figure}
  \centering
  \caption{The CNO cycle reaction. CREDIT: Wikipedia}
  \includegraphics[width=0.6\textwidth]
    {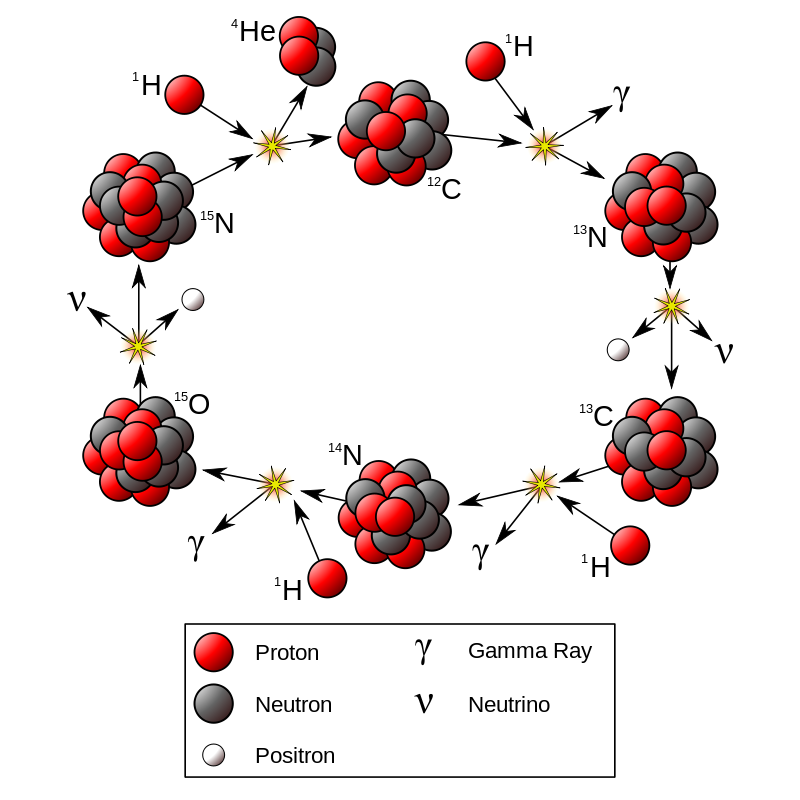}
  \label{cno}      
\end{figure}

\fcolorbox{gray}{DarkSeaGreen}{
\begin{minipage}[c]{.9\textwidth}
\begin{center}
\textbf{BOX 2: CNO CYCLE}
\end{center}

The Carbon-Nitrogen-Oxygen (CNO) cycle is the other set of fusion reactions that  convert hydrogen into helium in the stars.
Unlike the proton-proton chain reaction, CNO is a catalytic cycle. In  stars with mass larger than 1.3 solar masses, the CNO
cycle is the main source of energy according to theoretical models \citep{Salaris2005}. As illustrated in Figure \ref{cno}, 
four protons fuse, using isotopes of carbon, nitrogen and oxygen as catalysts, producing one alpha particle, two electron neutrinos
and two positrons. The same result is obtained in the CNO cycles, despite the different paths and catalysts involved. 

\label{box2}
\end{minipage}
}

\vspace{5mm}

\subsection{Stellar nucleosyntesis}
\label{subsec:3-3}

In his Nobel Prize Lecture, Hans Bethe said ``Stars have a life cycle much like animals. They are born, they grow, they go
through a definite internal development, and finally die, to give back the material of which they are made so that new stars 
may live" \citep{Bethe1968}.  Stars exist in the balance between two forces. On one hand, the star's gravity attempts to
compress the stellar material into the smallest possible sphere, and on the other hand, enormous pressure and heat are produced 
by nuclear reactions at the center of the star pushing all the material outward. The outcome of this balance depends largely on
the star's total mass. The stars are traditionally divided into three categories \citep{Rauscher2011}:

\begin{enumerate}
\item stars less massive than the Sun; 
\item stars with mass larger than the Sun and smaller than approximately eight solar masses; and 
\item stars more massive than eight solar masses. 
\end{enumerate}

The dominant reaction within small stars (in the first category) is the conversion of hydrogen into helium, whereas stars
in the second mass category have further reactions that convert helium to carbon and oxygen (see fact box 3). Only very 
massive stars in the third category support chain reactions that produce heavy elements up to the mass of iron.

\vspace{5mm}

\fcolorbox{gray}{BurlyWood}{
\begin{minipage}[c]{.9\textwidth}
\begin{center}
\textbf{BOX 3: NUCLEOSYNTHESIS OF CARBON AND OXYGEN}
\end{center}

When the fusion of protons into helium continues until the star has exhausted its hydrogen, the temperature in its core rises 
to about a few times 10$^8$ K, allowing the fusion of helium into heavier nuclei. In the first reaction two nuclei of helium, $^4$He, 
fuse with each other, creating the nucleus of  beryllium, $^8$Be. However, the $^8$Be nucleus has an extremely short mean life of 
just $10^{16}$ s, before it decays back again to two $^4$He nuclei. The rate of production equals the rate of destruction of $^8$Be nucleus:

\begin{equation}
^4He + ^4He  \leftrightarrow ^8Be
\label{eq:04}
\end{equation}

Nevertheless, the $^{8}$Be can capture another $^{4}$He nucleus producing the $^{12}$C nucleus by the reaction:

\begin{equation}
^8Be + ^4He  \longrightarrow  ^{12}C + \gamma
\label{eq:05}
\end{equation}

The reactions in Equations \ref{eq:04} and \ref{eq:05} are called the triple-alpha reaction, because three $^{4}$He nuclei or
alpha particles are necessary for the creation of $^{12}$C. This reaction can only create carbon in appreciable amounts because
of the existence of a resonance in $^{12}$C at the relevant energy for helium burning. Through this resonance the reaction in 
Equation \ref{eq:05} is enhanced by many orders of magnitude. 

The production of oxygen nuclei $^{16}$O is the result of a capture of another $^{4}$He nucleus by the carbon nuclei created in helium burning:

\begin{equation}
^{12}C + ^4He  \longrightarrow  ^{16}O + \gamma
\end{equation}

About half of the carbon nuclei produced are converted into oxygen.
\label{box3}
\end{minipage}
}

\vspace{5mm}

It is interesting to note that the elements $^{8}$C and $^{16}$C are extremely fine-tuned with respect to the nuclear force. In
the case of the strength of this force were just 0.5\% different from their current values, the average abundance of carbon or
oxygen in the universe would be more than two orders of magnitude smaller. That would make life based on carbon much more difficult
to occur \citep{Oberhummer2000, Schlattl2004}.

\subsection{Water molecule}
\label{subsec:3-4}

Once a star like the Sun has exhausted its nuclear fuel, its core collapses and the outer layers are expelled as a planetary nebula,
while the massive stars (more than eight solar masses) can explode in a supernova as their inert iron cores collapse. At these stages 
huge quantities of new nuclei are rapidly synthesized in the nuclear reactions triggered by the flood of neutrons. Most of the elements
heavier than the iron group are generated either by nucleosynthesis, or by radioactive decay of unstable isotopes that were produced. 
This material ejected at the end of life of such stars resulted in huge interstellar clouds of gas and dust. In general, the gas is made
of about 90\% hydrogen, 9\% helium, and 1\% heavier atoms, while the dust is composed of silicates, 
carbon, iron, water ice, me-thane, ammonia, and some organic molecules \citep{Dalgarno2006}. 

Therefore, the first water molecules of the universe might have been emerged in interstellar clouds produced at the end of life of the 
first generation of massive stars. Interstellar clouds are abundant in our galaxy, and it is generally considered that all stars and 
planets have been formed from them. 

\section{Distribution of water in the universe}
\label{sec:4}

\subsection{Water in galaxies}
\label{subsec:4-1}

 Water vapour in galaxies is best detected by observing maser emissions. Masers (Microwave Amplification by Stimulated Emission of Radiation) are similar to lasers, only emitting microwave radiation instead of visible light. Water molecules can absorb energy available around them in High Mass Star-forming Regions (HMSR) or near dying stars, and re-emit it as microwave radiation. Several water masers were found in our Milky Way galaxy \citep{walsh2008pilot}. \cite{mochizuki2009survey} studied water masers from young stellar objects (YSOs) in the outer regions of the galaxy.
 
 Water masers were also found in nearby galaxies \citep{darling2008ubiquitous, darling2016water}. \cite{braatz2008discovery} reported eight new sources of water maser emission in surveys conducted in nuclei of a hundred of galaxies. \cite{tak2015water} reviewed the presence of water in nearby galactic nuclei and galactic interstellar clouds and concluded that the emission of radiation is necessary to detect water in those sites.
 
\subsection{Water in stars and interstellar space}
\label{subsec:4-2}

\begin{figure}
  \centering
  \includegraphics[width=0.7\textwidth]
    {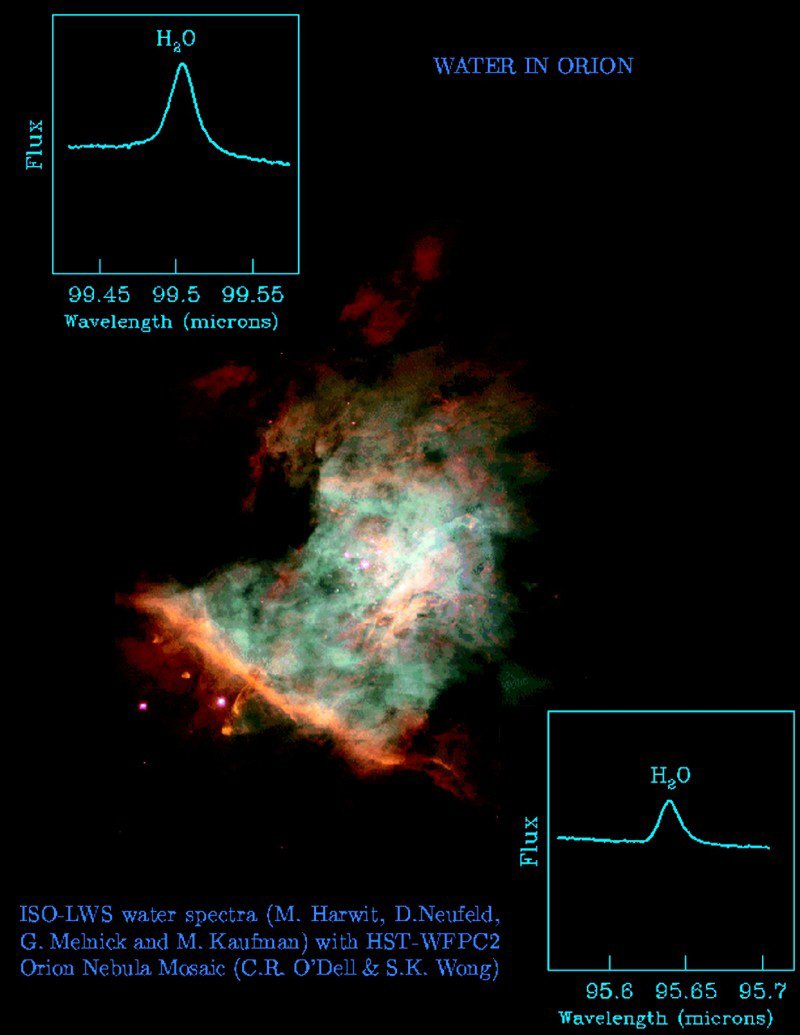}
  \caption{Water signatures in M42, the Orion nebula. Mosaic picture made from more than 40 individual Hubble Space Telescope (HST) images. Credit: ESA/NASA}
  \label{orionwater}       
\end{figure}

 The space between stars, the interstellar medium (ISM), is permeated with dust, gas in atomic, molecular and ionic forms, and cosmic rays. Stars are born out of dense regions within molecular clouds in the interstellar space and, when they die, the interstellar medium is enriched with elements heavier than helium (see Section \ref{sec:3}) \citep{hanslmeier2010water}. The Orion nebula is an example of star-forming regions (Fig. \ref{orionwater}).
 
Water ice is abundant in the interior of molecular clouds \citep{allamandola1988laboratory}. \cite{cheung1969detection} were the first to detect of water molecules in the interstellar medium. \cite{elitzur1989H2O} proposed a comprehensive model of water masers in star-forming regions. \cite{felli2007water} monitored a sample of 43 masers within star-forming regions for 20 years and created a database of their variability. More than 1010 galactic water maser sources have been listed in the Arcetri catalogue \citep{brand1994arcetri, valdettaro2001arcetri} and there is a distinction between water masers associated with star forming regions and late-type stars. \cite{furuya2003water} investigated water masers in low-mass young stellar objects.

 Mira was the first star, where water was detected in its spectrum, in 1963 \citep{kuiper1963}. \cite{russell1934molecules} had already predicted the presence of water in the atmosphere of late type stars\footnote{Stars cooler than the Sun (less than 5200 K), with a yellow-orange-red color variation}. He showed theoretically that water should be the most abundant molecule beside atomic or molecular hydrogen in stars of approximately 2800K. 

 Late type stars generally show strong mass loss and in many cases they form a circumstellar shell. \cite{maercker2008circumstellar} reported high water abundance in the circumstellar envelope of the star W Hya (an M-dwarf) and investigated water in the envelopes of six other M-type stars. They concluded that high amounts of water found in the majority of these sources may be explained by some kind of internal chemical processes. Updates on their research were reviewed by \cite{maercker2009circumstellar, maercker2016agb}

  \cite{jones2002spectral} analysed the spectra of a range of M stars and concluded that their observations match well with previous ground-based observations. 
  
  \cite{tsuji2000water} reported the presence of water in the spectrum of the M2 supergiant star $\mu$ Cephei. \cite{winnberg2008water} investigated the variability of water masers in circumstellar shells of late-type stars, using RX Bootis and SV Pegasi as representatives of semiregular variable stars (SRVs).

\subsection{Water in planetary disks}
\label{subsec:4-3}

 Most emerging stars(protostars) have a protoplanetary disk that forms from a molecular cloud. Water was present in the molecular cloud that gave birth to our solar system. The presence of water played an important role as the cloud settled to form the protoplanetary disk from which the planets were formed \citep{mottl2007water}.  
 
    T Tauri are young stars that are often immersed in large molecular clouds and have accretion disks around them. \cite{shiba1993detection} conducted a survey of 52 T Tauri stars and found water vapor in the disk of 17 of them. They measured the temperature of the water vapor, which appeared to be 2000 K. It is expected that the temperatures vary throughout the disk: water is likely solid in the outer regions of the disk and in gas phase in the inner regions where the temperature is higher than the evaporation point ($\sim$  150 K).
    
   The habitable zone can be predicted as the distance range from the central star, where water can be found in liquid phase on a planet surface. In the solar system, it has been calculated by \cite{kasting1993habitable} to be from 0.95 to 1.15 AU. Other estimates were reported by \cite{Rasool1970runaway, hart1979habitable, fogg1992estimate, spiegel2010generalized, abe2011habitable, kopparapu2013habitable} and \cite{way2014exploring}.
    
   The snowline is the distance from the protostar where water (and other volatiles) condensates into ice. The amount of water on the surface of terrestrial planets in the habitable zone depends on the snowline location, which is determined by the temperature profile of the disk, properties of the star, mass accretion rate, and size distribution of dust grains \citep{mulders2015snowline}. Mulders et al. calculated the snowline location in disks around different stars using estimates of mass accretion rates as a function of stellar mass. They used N-body simulations to predict the amount of water on the surface of terrestrial planets within the habitable zone. In addition, they determined that the variation of the snowline locations strongly affects the range of the water availability on terrestrial planets. They showed that a significant fraction of terrestrial planets in the habitable zone around Sun-like stars  remained dry (assuming ISM-like dust sizes) and no water was predicted on planets within the habitable zones of low-mass M stars ($<$ 0.5 M$_\odot$). When considering larger grains of dust, the snowline got closer to the star and that enabled water to be delivered to the habitable zone of a significant fraction of M stars and all FGK stars.

  A protoplanetary disk has its temperature and density profiles defined by the mass fraction of micrometre-sized dust grains and on their chemical composition \citep{bitsch2016influence}. The larger is the abundance of micrometre-sized grains, the higher is the overall temperature of the disk, and the further away from the star is the snowline. If the dust abundance is kept the same, an increase in the water fraction inside the disk may lower the temperature in the inner regions and increases the temperature in the outer regions of the disk. Disks with a smaller water fraction have the opposite effect. \cite{bitsch2016influence} studied the formation and migration of planets exploring the dust composition and its abundance in the disk.  Their results imply that hot and warm super-Earths may contain a significant fraction of water, if they have formed beyond the snowline and migrated inwards.
 
  Water vapor was found in the circumstellar disks such as AA Tauri \citep{hanslmeier2010water}, TW Hydrae \citep{salinas2016first}, and APM 08279+5255 \citep{lis2011discovery}. \cite{eisner2007mwc480} reported water within 1 AU of the young star MWC 480, which is believed to have resulted from the sublimation of inwards migrating icy bodies, that provided water for potential terrestrial planets. \cite{salyk2015detection} reported the presence of water vapor in the protoplanetary disk around DoAr 44.

\subsection{Water in extrasolar planets}
\label{subsec:4-4}

 By 2016, more than 3,300 extrasolar planets have been confirmed, with more than 570 multi-planet systems were reported \citep{nasa2016exoplanets}. Detecting atmospheres on extrasolar planets is a challenging task.
 
 The planet HD209458b was detected in 1999 and was the first extrasolar planet confirmed by the transit method. It is a giant planet with mass of approximately 63\% the mass of Jupiter. It is 100 times closer to the central star than Jupiter. With this proximity to the star, it is assumed that the planet continuously loose volatiles, and the outflow is estimated to be 10$^4$ tons/s. Water is among these volatiles as revealed by the data from Hubble Space telescope \citep{rauer2004search}.
 
  Planets with masses between 1 and 10 of Earth masses are known as super-Earths. The habitability of super-Earth planets was discussed, for example, by \cite{kaltenegger2008habitability}. It is likely that super-Earths have a wide range of atmospheres types. \cite{miller2009atmospheric} argued that some of them may have hydrogen-rich atmospheres.
  
  \cite{dominguez2016abundance} analysed the abundance of water and its dependency on stellar metallicity in extrasolar planetary systems. They found that the ratio of H$_2$O/SiO$_2$ produced in a molecular cloud of solar metallicity can account for  the ratio of these compounds on Earth today, supporting the ``wet'' hypothesis that implies that Earth could have obtained enough water locally during its formation \citep{stimpfl2004adsorption}.  \cite{bialy2015} studied the first step of H$_2$O formation in molecular clouds with extremely low metallicity, and showed that these clouds could have high abundances of water vapor.  Some of this water may have contributed to forming planets if the cloud collapses into a protoplanetary disk.
   
    \cite{ehrenreich2007spitzer} searched for water in the transit exoplanet HD189733b by using the Spitzer telescope and showed that the observational capabilities in that moment were insufficient for detecting water vapor. Water vapor was confirmed in the atmospheres of extrasolar planets HD 189733 b \citep{barman2008water, mccullough2014water}, HD 209458 b \citep{beaulieu2010water}, Tau Boötis b \citep{lockwood2014water}, HAT-P-11b \citep{fraine2014water, hanslmeier2010water}, XO-1b, WASP-12b, WASP-17b, and WASP-19b \citep{mandell2013exoplanet}.

\subsection{Water in the solar system}
\label{subsec:4-5}

 Water is very abundant in the solar system. It is present even in the Sun, as confirmed for example by \cite{wallace1995water}. 
 
 The solar system can be divided in two distinct group of planets by their position relative to the snowline: in the inner solar system, the volatiles are in a gaseous form and planets are relatively dry, small and rocky - the terrestrial planets, whereas in the outer solar system, the volatiles are in a solid form and planets are big, gaseous with a solid rocky-ice inner core - the giant planets.
 
\subsubsection{Water in the outer solar system}
\label{subsec:4-5-1}

 Water is an important constituent of the four giant planets - Jupiter, Saturn, Uranus, and Neptune \citep{stevenson1981behavior}. All these planets have similar structure, with a rocky-ice core and a huge gaseous layer consisting mainly of hydrogen and helium. 
 
 The abundance of water in Jupiter's atmosphere was studied by \cite{roos2004water}. They found that the O/H ratio in the atmosphere of Jupiter was comparable with the one in the sun.  Water in Jupiter was also reported by \cite{bjoraker1986abundance}. \citep{hueso2004three} studied water storms in the atmosphere of Jupiter and concluded that they may develop velocities of up to 150 m/s.
 
  Jupiter has four large satellites, known as the Galilean satellites: Io, Europa, Ganymede, and Callisto. Water keeps flowing away from Io \citep{pilcher1979stability}, which may be explained by the thermal escape. \cite{kumar1982atmospheres} investigated the atmospheres of Io and other Jupiter satellites and found that Europa, Ganymede, and Callisto may have oxygen atmospheres resulting from photolysis of water vapor.  Moreover, these three satellites may have internal oceans more than a hundred kilometers thick \citep{spohn2003oceans}. \cite{leitner2014ocean} developed a model for the analysis of oceans on Europa and Ganymede, and compared the results with the measured composition of brines on the surface of Europa. \cite{vance2014ganymede} analysed the influence of salinity on the internal structure of Ganymede and predicted that water ice may be present in the aqueous magnesium sulfate. They concluded that the stability of ice under high-pressure implies water-rock contact.
 
  Water in the deep atmosphere of Saturn was studied by \cite{visscher2005chemical}, who discussed chemical constraints for the abundance of water in the planet. Saturn has 60 confirmed moons. In 1997, water was detected in the atmosphere of Titan, the largest satellite of Saturn. The observed water abundance appeared to be four times lower than that in comets, suggesting that Titan's atmosphere was formed by outgassing from the interior rather than having a cometary origin\citep{coustenis1998evidence}. More recent data are in good agreement with these findings \citep{nixon2006water, bjoraker2008cassini}. \cite{raulin2008astrobiology} describes Titan as ``another world, with an active prebiotic-like chemistry, but in the absence of permanent liquid water on the surface: a natural laboratory for prebiotic-like chemistry''. \cite{dunaeva2013titan} built models of Titan's possible internal structure and predicted that Titan consists of the rock-iron core, rock-ice mantle and outer water-ice shell. 
  
  There is a water influx from the Saturnian rings that is caused by its satellite Enceladus \citep{mueller2006water}; this has been discussed earlier by \cite{connerney1984model}. This sixth largest moon of Saturn is mostly covered by clean fresh ice, being one of the most reflective bodies of the solar system. In 2004, Cassini detected water vapor and complex hydrocarbons emerging from the geologically active south-polar region of Enceladus \citep{spencer2006cassini}. \cite{tobie2008solid} showed that its particular location at the south pole and the magnitude of dissipation rate can only be explained by the models that assume a liquid water layer at a certain depth. \cite{ingersoll2010subsurface} affirmed that the existence of liquid water on Enceladus depends on the efficiency of subsurface heat transfer. \cite{iess2014gravity} studied the interior structure of Enceladus and its gravity field; their results suggest that the body deviates mildly from the hydrostatic equilibrium.

   \cite{atreya2006ocean} assessed the existence of an ocean of water-ammonia on Neptune and Uranus. They argued that the tropospheric cloud structure and the existence of a magnetic field must be maintained by a water-ammonia ionic ocean creating a dynamo action.
 
  The five main satellites of Uranus, Miranda, Ariel, Umbriel, Titania and Oberon, have weaker bands of water ice in their infrared spectra than those in the spectra of Saturn's icy moons and Galilean satellites. The difference may be explained by the presence of other ices (e.g, NH$_3$ and CH$_4$) besides water on their surfaces \citep{Encrenaz2007searching}. 
 
 So far, water has evaded detection on the dwarf planet Pluto based on Earth-bound observatories. \cite{cook2015search} analysed all data on Pluto from the Linear Etalon Imaging Spectral Array (LESIA: a component of the New Horizons spacecraft) searching for the presence of water. \cite{brown2000evidence} presented evidences of water ice on Charon, Pluto's Satellite. 
 
\subsubsection{Water in small bodies}
\label{subsec:4-5-2}

 In 2006, the International Astronomical Union defined the term `Small Solar System Body' (SSSB) as an object in the solar system that is not sufficiently massive to be a planet or a dwarf planet, and it is not a satellite. SSSBs are generally located in the main asteroid belt between Mars and Jupiter, in the Kuiper belt outside the orbit of Neptune, and in the Oort cloud extended as far as 50,000 AU from the Sun. Comets and asteroids consist mainly of pristine material and are remnants from the formation of the solar system about 4.6 billion years ago.
 
 \paragraph{Comets}  
 
  Comets are icy SSSBs that start a process of outgassing when approach the inner solar system. Comets  expel vaporized volatile materials, carrying dust away with them. They may have been an important source of water on Earth and other terrestrial planets, as well as on satellites. Water is the main component of interstellar and cometary ices \citep{allamandola1988laboratory}.
 
 Water in comets was first detected in 1970 from H and OH observations in comet Halley \citep{mumma1986detection, combes1988halley}. The comet Hale-Bopp (C/1995 O1) had its spectra analysed by
\cite{davies1997detection}, who found that ``some of the absorption features can be matched by an intimate mixture of water ice and a low-albedo material such as carbon on the nucleus''. \cite{cosmovici1998puzzling} detected water on comet Hyakutake (C/1996 B2). \cite{bock2009water} used  the Spitzer Space Telescope to detect water on the comets 71P/Clark and C/2004 B1 (Linear). \cite{schulz2006detection} detected water ice grains on Comet 9P/Tempel 1 by analysing the results of the DEEP IMPACT mission. \cite{bergh2004kuiper} presented general remarks about water ice and organics in the Kuiper belt objects.
 
 \cite{hsieh2006population} discovered comets in the main asteroid belt, a new class of objects in the solar system. The activity of these comets is consistent with dust ejection driven by water-ice sublimation. 
 
 \paragraph{Asteroids}  

 Asteroids and comets were previously thought to be of different origin: it was assumed that asteroids formed inside the orbit of Jupiter and comets originated from the outer solar system. However, the discovery of main-belt comets and recent findings like the returned sample of comet 81P/Wild 2 \citep{ishii2008comparison} have blurred the distinction between comets and asteroids. 
 
  Every year, thousands of new asteroids are found and several thousands of asteroids have been studied. The first evidence of water in an asteroid was found in Ceres, now classified as a dwarf planet. \cite{lebofsky1978ceres} estimated that Ceres's surface may contain 10\%$-$15\% water of hydration\footnote{ Hidration implies that water molecules are components of the crystal structure of a mineral. In this case a new mineral is created, a hydrate.}. \cite{kuppers2014ceres} indicated that at least 10$^{26}$ water molecules per second are being evaporated from the dwarf planet, and this phenomenon could be due to ``comet-like sublimation or cryo-volcanism, in which volcanoes erupt volatiles instead of molten rocks''. 
  
  \cite{fanale1989water} analysed the spectral signature of water in asteroids and confirmed that 66\% of the C-class asteroids in the investigated sample have hydrated silicate surfaces. Although it was believed  that D-type asteroids do not have water \citep{barucci1996water}, \cite{kanno2003first} suggested that these asteroids could contain water ice or hydrated minerals.
  
  \cite{yang2007spectroscopic} investigated spectral signatures of water ice on Trojan asteroids\footnote{ These asteroids are located near the equilibrium points L4 and L5 in the Sun-Jupiter system.} and their analysis showed no signs of water. \cite{treiman2004vesta} analysed the meteorite Serra de Magé, an eucrite believed to come from the asteroid 4 Vesta, and inferred that polar ice deposits in Vesta and similar asteroids are possible remainders from comet impacts, similar to water ice deposits on the Moon and Mercury. \cite{campins2009themis} used IR observations to confirm water ice on the surface of asteroid 24 Themis.
 
 \paragraph{Meteorites}
 
  A meteoroid is a small rocky or metallic body moving around the Sun or in the outer space. They are significantly smaller than asteroids or comets. Most meteoroids are fragments from asteroids or comets, other originated from debris ejected from impacts on bodies such as Mars or the Moon. A meteoroid that reaches the surface of the Earth without being completely vaporized is called meteorite. 
  
  \cite{ashworth1975water} analysed hydrous alteration products of olivine (a magnesium iron silicate) in an ordinary chondrite and an achondrite $-$ two classes of meteorites in which hydrous minerals are rare. Their observations suggest that both meteorites have unusual volatile constituents, and they argue that the Nakhla achondrite contains water of extraterrestrial origin, and this may also be the case for  the Weston chondrite,.
  
   The Shergotty$-$Nakhla$-$Chassigny meteorites, believed to be of martian origin, contain 0.04\%$-$0.4\% water by weight. \cite{karlsson1991water} used oxygen isotopic analysis to resolve whether this water was terrestrial or extraterrestrial. The results revealed that some of the water was extraterrestrial.

   Carbonaceous chondrites are a class of chondritic meteorites that includes the most primitive known meteorites. They contain high percentages of water (3\% to 22\%) \citep{norton2002cambridge} and organic compounds. Water and D/H ratios in the Chainpur (an ordinary chondrite) and Orgueil (a carbonaceous chondrite) meteorites were measured by \cite{robert1978water}. Orgueil is one of the most studied meteorites, owing to its unique primitive composition and relatively large mass \citep{gounelle2014orgueil}.

\subsubsection{Water on Earth and other terrestrial planets}
\label{subsec:4-5-3}

 Mercury, Venus, Earth, Mars and their satellites, have very different histories. Water is present in all these terrestrial planets, but in very distinct patterns: Mercury has ice at the poles; Earth has very abundant water, in lakes, oceans, underneath the surface, and in icy continents; Venus has vapor in the atmosphere; and Mars has ice at the poles, liquid water in salty flows, and vapor in the atmosphere.

 \paragraph{Mercury}
  Mercury, the smallest and closest planet to the Sun, has temperature of 450$^\circ$C on the dayside and  $-$180$^\circ$C on the nightside. This extreme contrast is a consequence of the lack of substantial atmosphere. \cite{butler2001nature} used a radar system to analyse Mercury surface and found that water could persist near the poles of Mercury inside of deep craters. \cite{wood1992temperatures} structured a thermal model that predicted the temperatures on the surface of Mercury and argued that, despite the proximity to the sun, temperatures at the poles could be low enough to permit water ice, as long as the albedo was high. Water was confirmed on Mercury by MESSENGER in 2012 \citep{lawrence2013evidence}.
 
 \paragraph{Venus}
  Venus is similar to Earth in its size and has a dry surface hidden by dense clouds. Contrary to Mercury, it has a dense atmosphere mainly consisting of the greenhouse gas CO$_2$, with a surface pressure of 90 times that of Earth's. The mean surface temperature is approximately 460$^\circ$C, and thus finding ice or liquid water near the poles of Venus is unlikely. Water vapor is an important component of the atmosphere and contributes to the global greenhouse effect. Water is mainly found below the cloud base at approximately 47 km above the surface \citep{hanslmeier2010water}. \cite{fedorova2008hdo} measured vertical distributions and mixing ratios of H$_2$O and HDO in Venus' mesosphere. They reported an increase of deuterium closer to the surface indicating a lower escape rate of D atoms comparing to H atoms or (and maybe also) a lower photodissociation of HDO comparing to H$_2$O. Water loss of Venus was measured by \cite{delva2009hydrogen} as 10$^{24}$ molecules per second. 
  
 \paragraph{Earth}
  Earth is the largest terrestrial planet, and the oceans comprise 2/3 of its surface. Between a very hot and a very cold planet lies the Earth, where the average surface temperature of 288 K and pressure of 1 bar create a favorable environment for life, and where water can be found as vapor, ice, and liquid, simultaneously. 
  
\begin{figure}[h]
   \caption{\textit{The different values of the deuterium-to-hydrogen ratio (D/H ratio) in water, observed in various bodies of the solar system. (Source: \cite{esa2017esa})}} 
   \includegraphics[width=1\columnwidth]{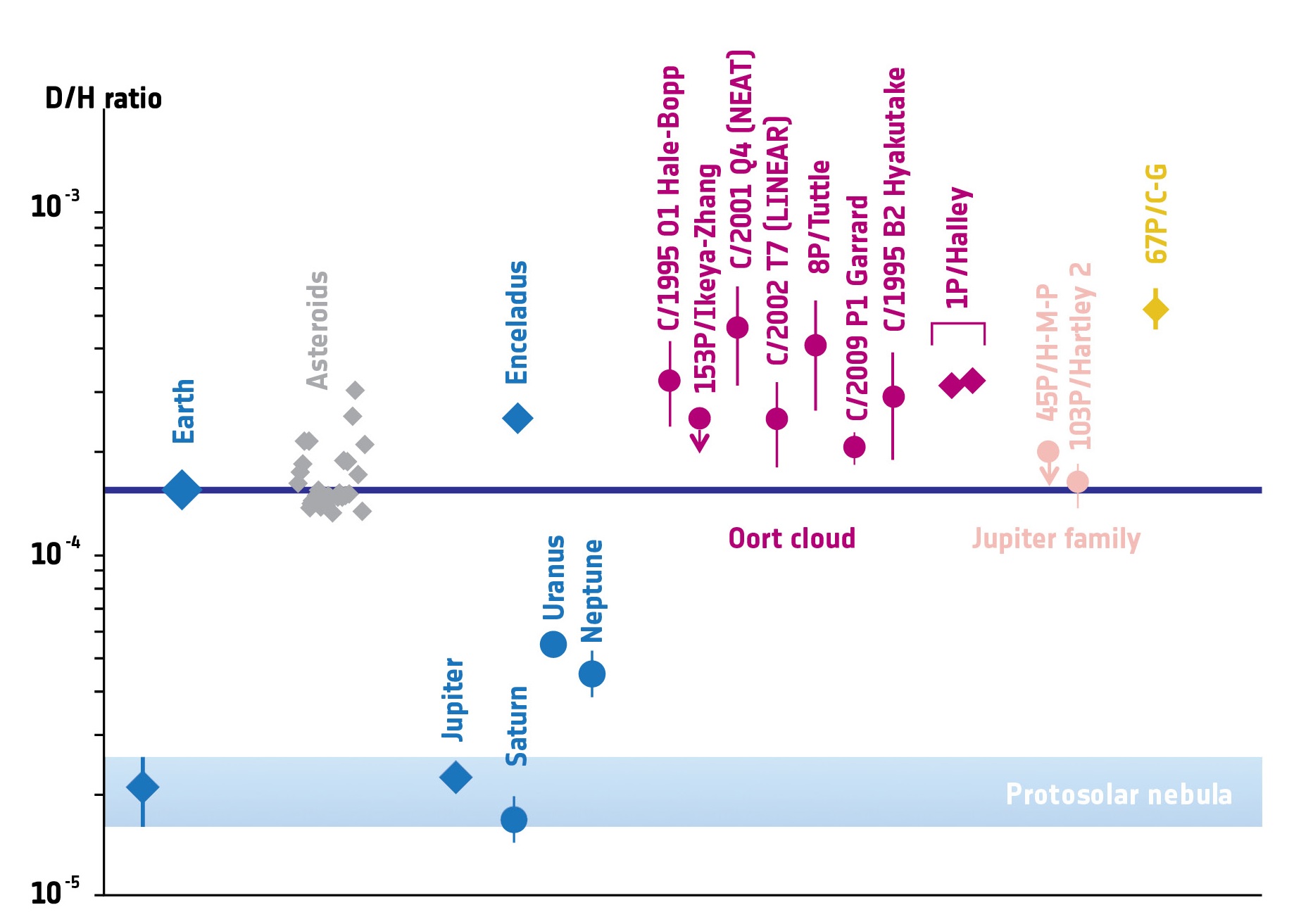}
   \label{figura:DHSolarSystem}
\end{figure}

  Water is present in large quantities in the Earth's atmosphere, along with 77\% of N$_2$, 22\% of O$_2$, and 1\% of other gases. An estimate of the amount of water inside Earth points to values ranging from 1 O$_\oplus$\footnote{O$_\oplus$ = mass of Earth’s oceans = 1.4 $\times$ 10$^{24}$ g} to 50 O$_\oplus$, with $\sim$10 O$_\oplus$ being the most likely value \citep{drake2006origin}. The origin of water on Earth is one of the most intriguing and debated issues in astronomy. One way to tackle the question is to analyse the proportion between the heavy water (HDO) and the light water (H$_2$O), also known as the D:H ratio, in Earth's water and in different potential sources. This ratio in the present day Earth mantle and oceans (Standard Mean Ocean Water, SMOW) is about 6 times higher than in the gas of the proto-planetary disk. Other bodies in the solar system present a great variation in their water D/H ratio, as shown in Fig. \ref{figura:DHSolarSystem}. Although the comparison of this ratio on Earth with those of various meteorite types suggests that the water on Earth was derived mainly from asteroids, the remnants of the protosolar nebula are still present in the Earth mantle, presumably signing the sequestration of nebula gas at an early stage of planetary growth \citep{Marty2012origins}. Marty has proposed that a small ($\leq$10\%) fraction of the mantle volatiles might have been derived from the protosolar nebula during an early stage of the proto-Earth growth. A small contribution, up to 10\%, may have come from comets \citep{morbidelli2000source}. \cite{izidoro2013compound} developed a model that considers all main possible sources of water and uses the D:H ratio to evaluate potential relative contributions from each source. 
  
  Regarding our satellite, \cite{sridharan2010moon} reported evidences of water ice at high latitudes on   the moon surface.
  
 \paragraph{Mars}
  Mars, the outmost terrestrial planet, has a mass of 0.1 M$_\oplus$ and surface temperatures in the range from $-$100$^\circ$C to 0$^\circ$C. Water ice is present in the two polar caps of the planet: it forms almost the entire north polar cap and the bottom layer of the south polar cap \citep{bibring2004perennial, christensen2006water}. Water vapor was detected in the atmosphere of Mars by \cite{owen1969mars}, and later quantified using Viking observations \citep{fedorova2010viking} and data from the Curiosity rover \cite{mahaffy2013abundance}.
  
\begin{figure}
  \centering
  \includegraphics[width=0.7\textwidth]
    {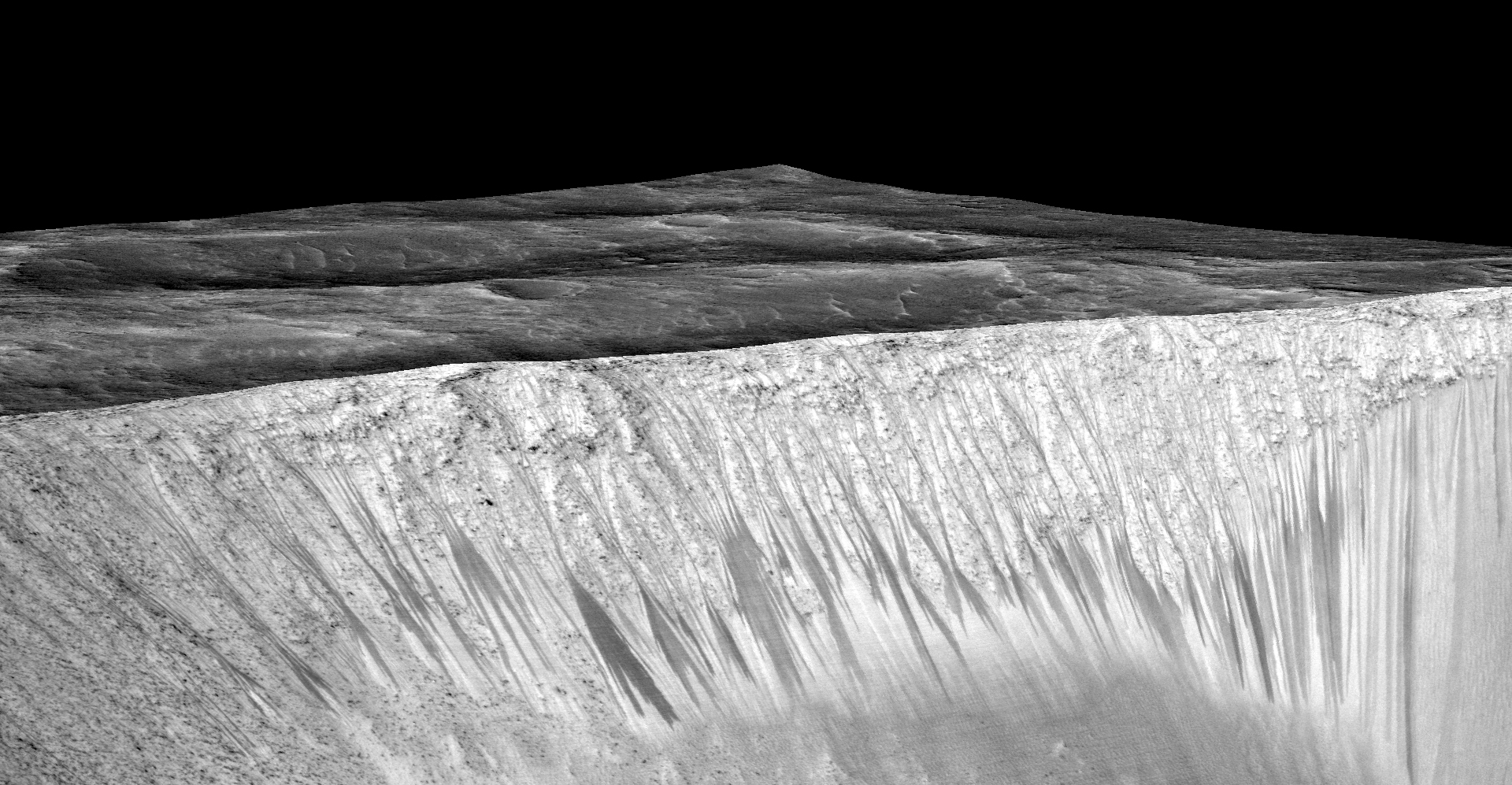}
   \caption{Garni crater on Mars with recurring slope lineae (dark narrow streaks) in its walls. The dark streaks are believed to be the flow of salty liquid water on Mars. Credit: NASA/JPL/University of Arizona}
  \label{watermars}      
\end{figure}

  \cite{masson2001geo} presented geomorphologic evidence that the planet underwent hydrologic cycle with liquid water on its surface in the past. \cite{ojha2015spectral} presented new evidences that salty liquid water flows sporadically on the present-day Mars despite of the low atmospheric pressure (1\% of that on Earth) and low temperature. Features known as recurring slope lineae (RSL), first identified in 2011, apparently resulted from the flows of salty liquid water on the surface of Mars. The look as dark streaks (Figure \ref{watermars}) and appear seasonally. The water remains in a liquid phase at low temperatures due to the presence of salt, which also protect the water from boiling off in the thin atmosphere of Mars. \cite{gough2016formation} analysed the formation of liquid water via the deliquescence of calcium chloride at low temperatures and concluded that calcium chloride may help to form liquid water that could cause slope streaks on Mars. \cite{pal2016possibiity} also predicted the appearance of microscopic amount liquid water on the hygroscopic mineral surfaces on Mars.

\section{Water and life}
\label{sec:5}

  It is not easy to define life in the astrobiology context. First there is no consensus on the time and mechanisms of the origin of life. Second, the distinction between living and non-living systems can get vague. Are viruses forms of life? Is there any kind of artificial life?
  
   Despite of this uncertainty, the established features of living organisms include: an organized structure to perform specific functions, including cells as fundamental units of life; performing of anabolic and catabolic reactions to sustain life (metabolism); regulation of internal conditions to keep them stable even in the unstable external environments (homeostasis); reaction to the environment stimuli or changes (response); growth; reproduction; and adaptation of organisms and populations to the environment (evolution) \citep{koshland2002seven}. 
  
  Other aspects of life like being carbon-based or having its genetical information in the form of DNA, may also be considered. Even then, all these characteristics define a single model of life as we know it. It is an arduous task to envision different forms of life that may exist elsewhere in the universe. That is why, so far, we only seek for established aspects the can provide conditions to support life as we know it. 

  Water is essential for life as we know it due to its unique properties. Because water can be found in all three phases on Earth, it allows an ample diversity of climates, habitats, and complex synergies between physical and chemical reactions \citep{schulze2004life}. It is an ideal solvent because of its polarity, and thus it can dissolute many different chemicals essential for metabolic reactions. Furthermore, the dipole character of water allows for hydrophobic organic molecules (e.g., lipids) to make cellular membranes. Alternate solvents may be possible, but there is a consensus that water is a prerequisite for life \citep{mottl2007water}.
  
   However, the presence of water does not imply life. So far we know very little about the probability of the emergence and evolution of life in a cosmic body that contains water. This uncertainty stems from the fact we still do not know how life began on Earth. Was it brought from outside, or may be Earth formation and evolution somehow made it possible for life to begin here? In this case, are those initial conditions replicable somewhere in the universe? The necessary chemical reactions and environmental conditions that allowed the emergence of life on Earth are still debated \citep{line2002enigma, trevors2004chance, pascal2006prebiotic, jortner2006conditions, benner2015martian}. At the present date, it remains impossible to determine precisely all the circumstances that led to the emergence of first living cells on Earth \citep{pascal2006prebiotic}.
   
   Nevertheless, one thing is certain: water must be in liquid state for all the living organisms we know. Although water is present everywhere in the universe (see Section \ref{sec:4}), liquid water seems to be extraordinarily rare. The water we have found so far on the surface of other cosmic bodies is always in the solid or gaseous states, but not liquid. The only exception we know so far (besides Earth itself) is the potential presence of small amounts of surface water on Mars. Direct measurement of temperature and pressure that could favour liquid water on remote planets and satellites is generally not possible today. Therefore, we are still far from mapping where life could emerge and evolve outside our planet \citep{Encrenaz2007searching}.
   
    In the last two decades, the astrochemical and astrophysical conditions for the emergence and evolution of life have been intensively debated (e.g., \cite{ehrenfreund2002origin} and \cite{chyba2005astrobiology}). Water on Earth, in the solar system and in the interstellar medium, and its strong association with life, has been reviewed by \cite{mottl2007water}. \cite{cottin2015astrobiology} presented an interdisciplinary review of astrobiology, covering the most recent facts and hypotheses.
    
   Even though we still have no evidence of the extraterrestrial life, the resilience and presence of life in a wide variety of environments on Earth, even in very challenging niches, suggest that life may not be restricted to our planet. Since the discovery of extremophilic microbes \citep{rothschild2001life, rampelotto2010resistance}, there has been less scepticism regarding the possibility of extraterrestrial life \citep{sagan1996circumstellar, chyba1997life, montmerle2006life}. Essentially, there are no chemical or physical barriers to extremophiles: where there is liquid water, there is life on Earth \citep{rothschild2001life}. Therefore, all the recent discoveries, including liquid water on Mars \citep{ojha2015spectral}, the possibility of liquid oceans underneath the surface of Europa and Enceladus \citep{raulin2005exo, tobie2008solid}, and also the presence of organic molecules on Titan \citep{raulin2008astrobiology}, have fueled astrobiological interest in the solar system and beyond. 
 
 Finding extraterrestrial life seems to be only a question of time now. As we have seen in this chapter, water is a key aspect in this search. That is why understanding how water came to be and spread in the universe give us the first steps in this great endeavour of looking for life outside of our planet. Once we succeed, it will certainly expand our knowledge about what is to be a living organism in this vast universe. Once knowing we are not alone, human beings will never look at themselves at the same way again.
 
\paragraph{Acknowledgements}
This work was partially funded by CNPq and FAPESP (proc. 2011/08171-3). This support is gratefully acknowledged. The authors also would like to thanks Tais Ribeiro for making Figure 1.

\section{Acronyms}

HMSR - High Mass Star-forming Regions

\noindent HST - Hubble Space Telescope

\noindent ISM - Interstellar Medium

\noindent LESIA - Linear Etalon Imaging Spectral Array

\noindent Masers - Microwave Amplification by Stimulated Emission of Radiation

\noindent RSL - Recurring Slope Lineae

\noindent SMOW - Standard Mean Ocean Water,

\noindent SRVs - Semiregular Variable Stars

\noindent SSSB - Small Solar System Body

\noindent YSOs - Young Stellar Objects


\bibliographystyle{apalike}   
\bibliography{bibliography}

\end{document}